\documentclass[
pra, 
floatfix,%
 reprint,
superscriptaddress,
nofootinbib,
 amsmath,amssymb,
 aps,
]{revtex4-2}

\usepackage{graphicx}
\usepackage{dcolumn}
\usepackage{bm}
\usepackage{hyperref}

\usepackage{multirow} 
\usepackage{xcolor, booktabs}

\usepackage{subfigure}
\usepackage[linesnumbered,ruled,vlined]{algorithm2e} 

\usepackage[normalem]{ulem}
\usepackage[capitalise]{cleveref}

\begin{document}

\preprint{APS/123-QED}

\title{Coincident Learning for Beam-based RF Station Fault Identification Using Phase Information at the SLAC Linac Coherent Light Source}

\author{Jia Liang} 
\email{jialiang@slac.stanford.edu}
\affiliation{Institute for Computational and Mathematical Engineering, \\ Stanford University, Stanford, California 94305, USA} 

\author{William Colocho}
\affiliation{SLAC National Laboratory, Menlo Park, California 94025, USA} 

\author{ Franz-Josef Decker}
\affiliation{SLAC National Laboratory, Menlo Park, California 94025, USA} 

\author{Ryan Humble} 
\affiliation{Cerebras System, Sunnyvale, California 94085, USA }

\author{Ben Morris}
\affiliation{SLAC National Laboratory, Menlo Park, California 94025, USA} 

\author{Finn H. O'Shea}
\affiliation{SLAC National Laboratory, Menlo Park, California 94025, USA} 

\author{David A. Steele} 
\affiliation{SLAC National Laboratory, Menlo Park, California 94025, USA} 

\author{Zhe Zhang} 
\affiliation{SLAC National Laboratory, Menlo Park, California 94025, USA} 

\author{Eric Darve} 
\thanks{Also at Mechanical Engineering, Stanford University, Stanford, California 94305, USA} 
\affiliation{Institute for Computational and Mathematical Engineering, \\ Stanford University, Stanford, California 94305, USA}

\author{Daniel Ratner} 
\affiliation{SLAC National Laboratory, Menlo Park, California 94025, USA}

\date{May 21, 2025}

\begin{abstract}

Anomalies in radio-frequency (RF) stations can result in unplanned downtime and performance degradation in linear accelerators such as SLAC’s Linac Coherent Light Source (LCLS). Detecting these anomalies is challenging due to the complexity of accelerator systems, high data volume, and scarcity of labeled fault data. Prior work identified faults using beam-based detection, combining  RF amplitude and beam-position monitor data. Due to the simplicity of the RF amplitude data, classical methods are sufficient to identify faults, but the recall is constrained by the low-frequency and asynchronous characteristics of the data. In this work, we leverage high-frequency, time-synchronous RF phase data to enhance anomaly detection in the LCLS accelerator. Due to the complexity of phase data, classical methods fail, and we instead train deep neural networks within the Coincident Anomaly Detection (CoAD) framework. We find that applying CoAD to phase data detects nearly three times as many anomalies as when applied to amplitude data, while achieving broader coverage across RF stations. Furthermore, the rich structure of phase data enables us to cluster anomalies into distinct physical categories. Through the integration of auxiliary system status bits, we link clusters to specific fault signatures, providing additional granularity for uncovering the root cause of faults. We also investigate interpretability via Shapley values, confirming that the learned models focus on the most informative regions of the data and providing insight for cases where the model makes mistakes. This work demonstrates that phase-based anomaly detection for RF stations improves both diagnostic coverage and root cause analysis in accelerator systems and that deep neural networks are essential for effective analysis. 

\end{abstract}

\maketitle



\section{introduction}

Identifying and resolving anomalies in particle accelerators can improve both uptime and performance for users. The Linac Coherent Light Source (LCLS) experiences unplanned downtime and unexpected behavior, reducing beam stability and introducing noise sources that can interfere with user data analysis.
Among the various failure modes of LCLS, our primary concern is pinpointing 
anomalous behavior in the 82 radio-frequency (RF) stations that power the X-ray laser’s accelerator. Anomalies in the RF stations are an interesting source of study for two reasons \cite{humble2022beam}: First, they are a common failure mode, with several drops in RF station performance per hour during beam operation. Second, they have a significant impact on performance because faults directly affect the beam energy, causing fluctuations in the pulse energy and photon energy of the free-electron laser (FEL) delivered to users.  

The increasing complexity of particle accelerators has made it difficult for humans to monitor and manage faults effectively. For example, LCLS's control system manages 200,000 process variables, highlighting the need for data science techniques to enhance the automation of accelerator operations. To alleviate the burden on human operators, our previous work \cite{humble2022beam} proposed a beam-based method to identify changes in accelerator status by observing shot-to-shot data from the beam position monitoring system. By comparing these beam-based anomalies with data from individual RF stations, we can determine which RF station was responsible for the change in beam behavior. The classical statistical method from \cite{humble2022beam} can be effective, but requires significant effort and manual data inspection to choose the right hyperparameters. Shifts in data distribution over time can make the selected hyperparameters—and, consequently, the algorithm—ineffective. 
More critically, the classical approach has only been applied to the low-rate amplitude data stream. The full-rate phase data stream is expected to include far more information about anomalies, but is too complex to be analyzed by classical methods. Consequently phase data was not used in \cite{humble2022beam}.

Comparing to amplitude data, we find that working with phase data offers several advantages. First, the phase data captures new information, enabling detection of anomalies that amplitude-based detection misses. Second, the time-synchronous phase data provides critical insights, allowing us to distinguish whether a change in the accelerator status occurs simultaneously with an RF station change or in response to a previous issue. Third, the rich information contained in the phase data facilitates clustering of anomalies into distinct categories, each with unique signatures. This categorization can aid in the expert diagnosis of the root cause within the RF stations and expedite recovery. 

In our previous work \cite{humble2024coincident}, we found that Coincident Anomaly Detection (CoAD) is an effective approach to detect RF faults. CoAD is designed to identify anomalies in multi-modal systems by leveraging coincident behavior across two data slices, e.g., an individual subsystem's performance and an overall system output `quality.' Any subsystem issue should manifest as coincident anomalies in both the subsystem and the larger system output, and this behavior can be exploited to train deep learning anomaly detection algorithms \cite{humble2024coincident}. In the context of the RF particle accelerator task, for example, one data stream comes from a radio-frequency (RF) station subsystem (\(s\)), while another data stream originates from beam-position monitors (BPMs), which provide data on the final electron beam energy and quality (\(q\)). Our previous work demonstrated the effectiveness of CoAD in real-world datasets. 

The previous studies on RF stations produced reasonable results using time-asynchronous amplitude data, but ignored the richer information from time-synchronous phase data, which, due to its complexity, could not be analyzed by the statistical method from \cite{humble2022beam}. In this paper, we extend the earlier work by training deep neural networks to analyze the phase data. 

This paper makes three primary contributions. First, we present the first automated framework for analyzing RF phase data, enabled by deep neural networks (DNNs), which eliminates the need for manual inspection and facilitates scalable anomaly detection in particle accelerators. Second, we apply our method to operational data from the Linac Coherent Light Source (LCLS) and demonstrate a substantial improvement in detecting RF anomalies, significantly outperforming existing baseline approaches. Third, we show that phase signatures exhibit distinct clustering patterns that align with known fault types, providing additional diagnostic insights and enhancing the interpretability of the anomalies by revealing potential root causes. The rest of the paper is divided into the following sections: \cref{sec:related_work} introduces the relevant background and related work; \cref{sec:rf_anomalies} introduces RF anomalies and compares phase data with amplitude data. \cref{sec:method} presents the proposed methodology; \cref{sec:results} discusses the experimental results; and \cref{sec:conclusion} concludes the paper and outlines future work.

\section{Related Work}
\label{sec:related_work}

Machine learning (ML) techniques have been extensively applied to fault detection in particle accelerators. For instance, \cite{chen2024anomaly} introduced an unsupervised LSTM autoencoder-based anomaly detection framework for the CEBAF accelerator's orbit lock system, which achieved high fault detection rates by modeling normal operational patterns to detect deviations. Similarly, \cite{tennant2023smart} developed a Smart Alarm system for the CEBAF injector beamline based on a neural network inverse model that predicts machine settings from sensor readings. This system detected 83\% more anomalous conditions than the existing rule-based system and achieved 94.6\% accuracy in identifying the root cause of faults through discrepancies between predicted and observed settings.

Recent works \cite{marcato2023time, blokland2022uncertainty, lobach2024long, suetsugu2024machine, rajput2024robust, radaideh2023early, rahman2024accelerating, jiqing2024particle, yang2025few, yucesan2024machine, edelen2021anomaly} in this domain tend to focus on \textit{early fault detection} and \textit{fault prevention}, aiming to issue alerts \textit{prior} to the actual onset of faults to enhance machine protection and reduce operational downtime. For example, \cite{rahman2024accelerating} proposed a deep learning architecture combining LSTM and CNN to predict slowly developing faults in accelerating cavities at CEBAF, achieving 80\% accuracy up to one second before fault occurrence. Their model leverages multivariate RF signal data and enhances predictive precision through a consecutive window criterion and confidence thresholding. In a similar vein, \cite{marcato2023time} demonstrated the utility of temporal deep learning models such as LSTM and Temporal Convolutional Networks (TCNs) for identifying pre-fault signatures in RF control systems.

Most aforementioned approaches \cite{marcato2023time, blokland2022uncertainty, suetsugu2024machine, rajput2024robust, radaideh2023early, rahman2024accelerating, jiqing2024particle, yang2025few} rely on supervised learning, which necessitates labeled datasets. A subset of recent works ---such as those proposed in \cite{lobach2024long, chen2024anomaly, tennant2023smart, yucesan2024machine, edelen2021anomaly}---explores unsupervised strategies, but these  typically assume that the training data is entirely normal. For example, \cite{chen2024anomaly} had to manually exclude faulty data during training to uphold this assumption. However, a key challenge in the fault detection for RF stations is the lack of labeled fault data. In such settings, unsupervised anomaly detection methods are a natural choice, yet their effectiveness is often undermined when the training data is contaminated with unidentified anomalies or noise—a common scenario in particle accelerator environments. As a result, their performance can degrade significantly in practice. In contrast, the method presented in this work requires neither labeled data nor a purely normal training set. Our approach is robust to significant anomaly contamination in the training dataset.

Another emerging direction in this area is the integration of advanced ML techniques to enhance \textit{performance}, \textit{adaptability in low-data regimes}, and \textit{model robustness}. Regarding performance optimization, \cite{yucesan2024machine} employed neural architecture search and hyperparameter tuning, resulting in an 8\% increase in the true positive rate for errant beam detection. For adaptability in data-scarce scenarios, \cite{yang2025few} introduced a Bidirectional Discriminative Prototype Network (BiDPN) for few-shot fault diagnosis in particle accelerator power systems. Their approach trains solely on normal operational data and excels in multi-class fault diagnosis, outperforming conventional deep learning models under limited fault sample conditions. For model robustness, \cite{blokland2022uncertainty} proposed an uncertainty-aware Siamese Neural Network to predict errant beam pulses at the Oak Ridge Spallation Neutron Source, achieving low false positive rates while also providing calibrated uncertainty estimates. Similarly, \cite{rahman2024accelerating} employed Monte Carlo Dropout (MCDO) at inference time to quantify predictive uncertainty and improve model reliability. In our work, we leverage a CoAD framework to reduce uncertainty, requiring agreement across multiple data streams before confirming an anomaly. 

Methodologically, the works of \cite{blokland2022uncertainty} and \cite{rajput2024robust} bear similarities to our proposed CoAD framework. Both employ Siamese networks, which, like CoAD, utilize dual input streams. However, key differences distinguish our approach. Siamese networks require labeled input pairs---typically ``normal-normal'' or ``normal-abnormal''---whereas CoAD operates without any labeled data. Furthermore, Siamese networks typically use inputs from the same modality (e.g., the same sensor type), while CoAD is designed for multimodal learning, requiring inputs from heterogeneous sources, such as system-level beam data and subsystem-specific diagnostic signals.  Moreover, CoAD addresses a unique and critical challenge. In complex systems like particle accelerators, diagnostic signals are often noisy and contain numerous anomalies, most of which are benign or unrelated to true faults. CoAD is specifically designed to disentangle these benign anomalies from fault-indicative anomalies. Unlike conventional anomaly detection methods, which merely identify deviations from normal behavior, CoAD leverages coincident learning to differentiate anomalies that genuinely signal faults. This capability is essential for fault diagnosis in such environments, where simply detecting anomalies is insufficient.

\section{RF Anomalies}
\label{sec:rf_anomalies}

The performance of LCLS is evaluated by the quality of the X-ray beam it delivers, measured through factors including pulse energy, photon energy stability, and bandwidth. As in our previous work \cite{humble2022beam}, an anomaly is defined as any unforeseen change in the accelerator condition that ranges from diminishing the quality of the beam to a complete loss of beam. Among the various failure modes, we focus on identifying anomalies in the 82 RF stations because they power the X-ray laser's accelerator, directly impact beam energy, and represent a significant source of failures. 

\subsection{Amplitude data}
\label{sec:ampl_data}

The existing method leverages two sources of data: RF station diagnostics, which provide insight into fault origins, and beam-based diagnostics, which capture the pulse-by-pulse performance of the accelerator. 

The previous work used two RF station diagnostics: the real-valued amplitude signals (AMPL), or the summary amplitude mean out of tolerance status bit (AMM) which indicates whether the RF station's amplitude is outside a predefined range. 
Both the AMM and AMPL data are recorded asynchronously with the beam data at a rate of about 0.2 Hz and can lag behind the beam data by several seconds. 

The beam-based diagnostic data originates from stripline beam position monitors (BPMs) \cite{straumann2007lcls} located in dispersive regions, which are sensitive to fluctuations in beam energy. The BPM data are recorded synchronously at the full LCLS operational rate of 120 Hz. 
The accelerator contains 175 BPMs, each measuring the beam's transverse position (X, Y), and charge (TMIT). Among these 175 BPMs, our work focuses on four specific BPMs in dispersive regions: BPMS:LTUH:250, BPMS:LTUH:450, BPMS:DMPH:502, and BPMS:DMPH:693. For each selected BPM, we use its charge measurement along with either the X or Y position (depending on the dispersive direction), as shown in \cref{fig:ampl_vs_phase}. Previous work in \cite{humble2022beam} and \cite{humble2024coincident} cross-reference BPM data with RF AMM/AMPL data to detect RF anomalies. 

\subsection{Phase data} 

\begin{figure}[htbp] 
    \centering 
    \includegraphics[width=0.5\textwidth]{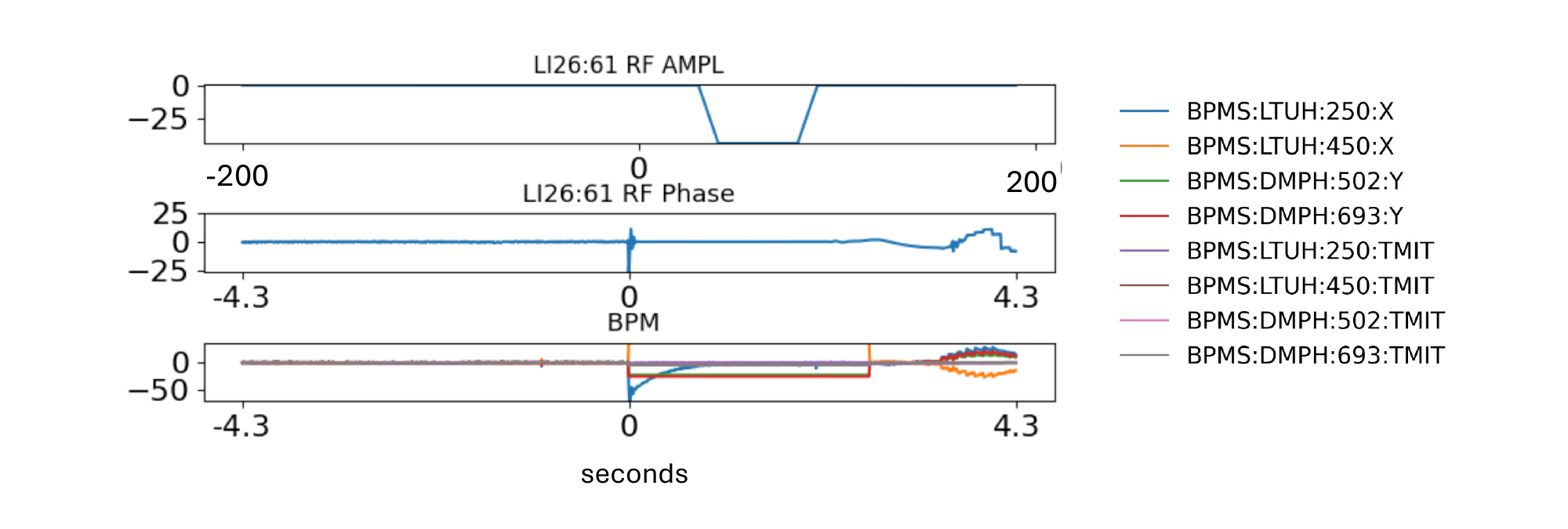} 
    \caption{Comparison of RF amplitude and RF phase diagnostic data during an RF fault in station LI26:61. The fault impacted downstream beam quality, as evidenced by the Beam Position Monitor (BPM) signals. BPM and RF Phase data is synchronous, and the data reflects correlated behavior. The RF amplitude diagnostic is slow, reporting only a step function change, and asynchronous with a multi-second lag relative to the phase and BPM data.} 
    \label{fig:ampl_vs_phase} 
\end{figure}

In this work, we aim to leverage phase data from the RF station subsystem(s) to detect RF anomalies. To ensure stable acceleration, the RF oscillations must align precisely with the periodic arrival of the particle beam in the cavity. The particle's arrival time relative to the RF cycle is referred to as the RF phase. Compared to amplitude data, phase data offers several advantages. First, the phase data is recorded at the full LCLS operation rate of 120 Hz, which matches the sampling rate of the BPM data. Moreover, the phase data is synchronized with the BPM data. A comparison between the phase and amplitude data is summarized in \cref{fig:amplitude vs phase data}, and an example of RF phase and RF amplitude can be seen in \cref{fig:ampl_vs_phase}. The figure illustrates a clear relationship between variations in phase data and corresponding changes in BPM behavior. Notably, at the 0-second mark, a distinct jump in the phase data aligns with a corresponding jump in the BPM. Toward the end of the observation window, the curved behavior of the RF phase also resembles the curved patterns observed in the BPM. In contrast, jumps in the AMPL data can occur more than 5 seconds apart from those in the BPM. 

\begin{table}[htbp]
\centering
\begin{tabular}{lll}
\toprule
              & Prior work \cite{humble2022beam, humble2024coincident}                                                              & Current                                                         \\
\midrule
Data Streams  & \begin{tabular}[c]{@{}l@{}}Amplitude (s) \\ BPM (q)\end{tabular}      & \begin{tabular}[c]{@{}l@{}}Phase (s) \\ BPM (q)\end{tabular}    \\
\midrule
Sampling Rate & \begin{tabular}[c]{@{}l@{}}s ($\sim$0.2 Hz)\\ q (120 Hz)\end{tabular} & \begin{tabular}[c]{@{}l@{}}s (120 Hz)\\ q (120 Hz)\end{tabular} \\
\midrule
Synchronicity & \begin{tabular}[c]{@{}l@{}}s lags behind q \space \space \\ up to 5 seconds\end{tabular} &  \begin{tabular}[c]{@{}l@{}} s and q are \\ synchronous \end{tabular} \\
\bottomrule
\end{tabular}
\caption{Comparison between RF phase diagnostic data and RF amplitude diagnostic data.}
\label{fig:amplitude vs phase data}
\end{table}

\section{Method} 
\label{sec:method}

\begin{figure}[htbp] 
    \centering 
    \includegraphics[width=0.4\textwidth]{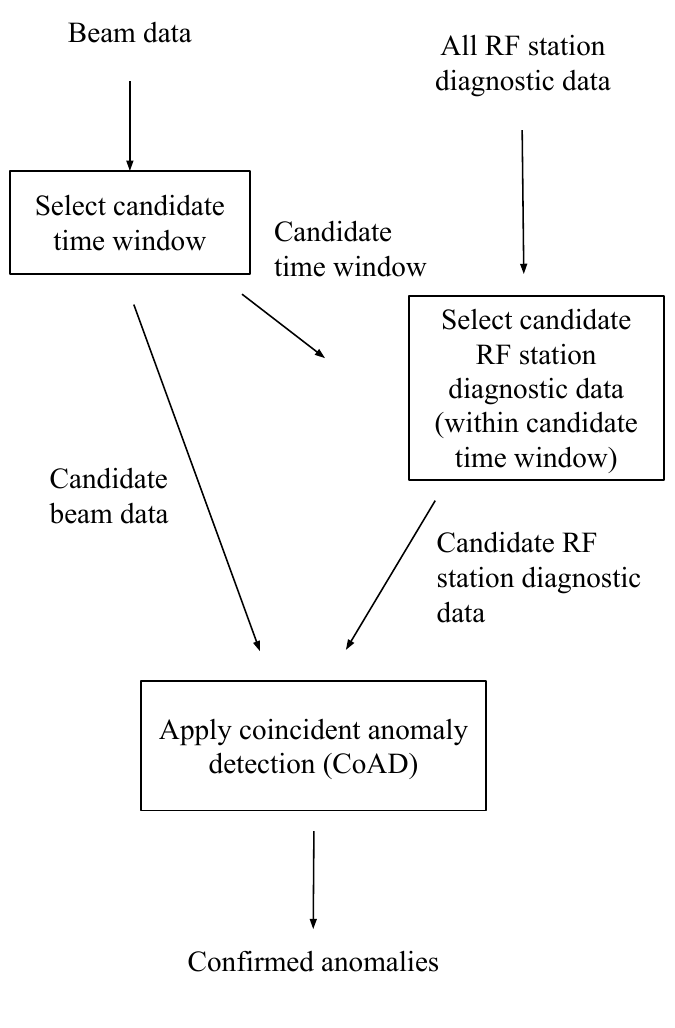} 
    \caption{Anomaly detection workflow: beam data is used to identify candidate time windows with potential anomalies, and RF diagnostic data highlights suspect RF stations. CoAD is then applied to detect anomalous behavior within the candidate windows.} 
    \label{fig:two-stage-method} 
\end{figure}

Our method consists of two main steps: `candidate' generation and CoAD.  The candidate generation stage reduces the data rate and enhances the anomalous fraction of the dataset fed to CoAD. Candidate generation itself involves two sub-steps: we first use beam data to identify time windows with potentially anomalous behavior, and then use RF station diagnostic data to identify potentially anomalous RF stations within each candidate window. For each such window, we pair the corresponding beam data with the diagnostic data from the selected RF stations to form anomaly candidates. CoAD is then applied to these selected candidates to identify true anomalies. This process is illustrated in \cref{fig:two-stage-method}. 

\subsection{Anomaly candidate generation}
\label{sec:anomaly_candidate_identification}

The candidate generation step uses beam data and RF stations' diagnostic data to generate anomaly candidates. The beam data is the multivariate time-series BPM data $x_{s,t}$ for signal $s$ at time $t$. The anomaly score is defined as $a_{s,t} = MAD(x_{s,t-2l}, ..., x_{s,t})$ where $l$ is the lagging rolling window size for computing the median and median-absolute deviation (MAD). Then, we aggregate the per-signal anomaly scores into a single anomaly score \( a_t \) using the geometric mean:
\[
a_t = \Big( \prod_{s=1}^{n} a_{s,t} \Big)^{\frac{1}{n}},
\]
where \( n \) is the number of signals. We further aggregate across $k$ consecutive pulses to get the final score 
\[
a_{AGG,t} = \Big(\prod_{i=0}^{k-1} \ a_{t-i} \Big)^{1/k} 
\]
to boost the score for sustained beam anomalies. The parameter $k$ controls the emphasis on sustained anomalies; higher values of $k$ prioritize longer-duration anomalies. 

We identify a candidate \( C_i \) centered at time \( t_i \) if the aggregated anomaly score \( a_{AGG, t_i} \) meets or exceeds a predefined threshold \( \tau \), i.e., \( a_{AGG, t_i} \geq \tau \). However, since the aggregated score \( a_{AGG, t} \) is compared with the threshold \( \tau \), the time point \( t_i \) satisfying \( a_{AGG, t_i} \geq \tau \) may not correspond to the earliest occurrence of the anomaly. To address this, we employ a simple heuristic to determine the earliest time point \( t'_i \) within the sequence where the anomaly first occurs. Specifically, we identify $t
'_i$ as the first point within the range $(t_{i-k}, t_i)$ whose anomaly score, $a_{t'}$, deviates by at least 1.25 standard deviations from the mean of the anomaly scores of all points in that range. We then use \( t'_i \) as the center of the anomaly candidate \( C_i \).

Once the center \( t'_i \) is determined, the full data window is defined as the range \( (t'_{i-w}, t'_{i+w}) \), where $w$ is the window size. In this work, the window size is set to approximately 530 samples, corresponding to roughly 4.3 seconds. This duration is chosen because beam faults can persist for several seconds, and recovery from an initial fault may also take a few seconds. Extending the window further does not provide significant additional benefits. The candidate \( C_i \) consists of two components. The first component is the beam data, which includes eight channels of BPM data as described in \cref{sec:ampl_data}. The second component is the RF diagnostic data, specifically the phase data. In our approach, we use the phase data from the RF station(s) exhibiting the most anomalous behavior around the center of the time window to identify the most likely root cause of the anomalies. In this work, the most anomalous RF stations are identified by measuring how much their phase values deviate from zero within a short time window preceding the beam trigger. The algorithm selects the five stations with the largest deviations, retains only those exceeding a predefined threshold, and pairs them with beam data for further analysis. If none exceeds the threshold, the station with the highest deviation is chosen. Additional details on the identification procedure are provided in ~\cref{sec:most_anomalous_rf_station}. Formally, let us denote the RF diagnostic data as \( s \), describing the subsystem, and the BPM data as \( q \), representing the overall quality of the system. Consequently, each candidate \( C_i \) is defined as \( C_i = \{s_i, q_i\} \).

\subsection{Coincident anomaly detection} 

Coincident Anomaly Detection (CoAD) \cite{humble2024coincident} is an unsupervised anomaly detection approach designed for datasets with paired inputs \( D = \{(s, q)\} \), where $s$ and $q$ are two streams of data. In the context of RF station fault detection, $s$ represents diagnostic data from the RF station subsystem, and $q$ represents a system-level quality indicator, such as measurements from beam position monitors. CoAD addresses tasks where fluctuations in $s$ and $q$ are uncorrelated during normal operating conditions, but both data streams are expected to change simultaneously during an anomalous event. By leveraging this coincidence principle, CoAD overcomes challenges faced by existing anomaly detection techniques, such as requiring labeled datasets or datasets known to contain only normal behavior.

CoAD is particularly useful in multimodal systems, where data comes from two distinct but related sources—such as different sensors monitoring the same system or different stages of a sequential process. Furthermore, the algorithm remains applicable even with a single measurement, provided that two inputs, $s$ and $q$, can be derived by partitioning the data.

\begin{figure}[htbp] 
    \centering  
    \includegraphics[width=0.5\textwidth]{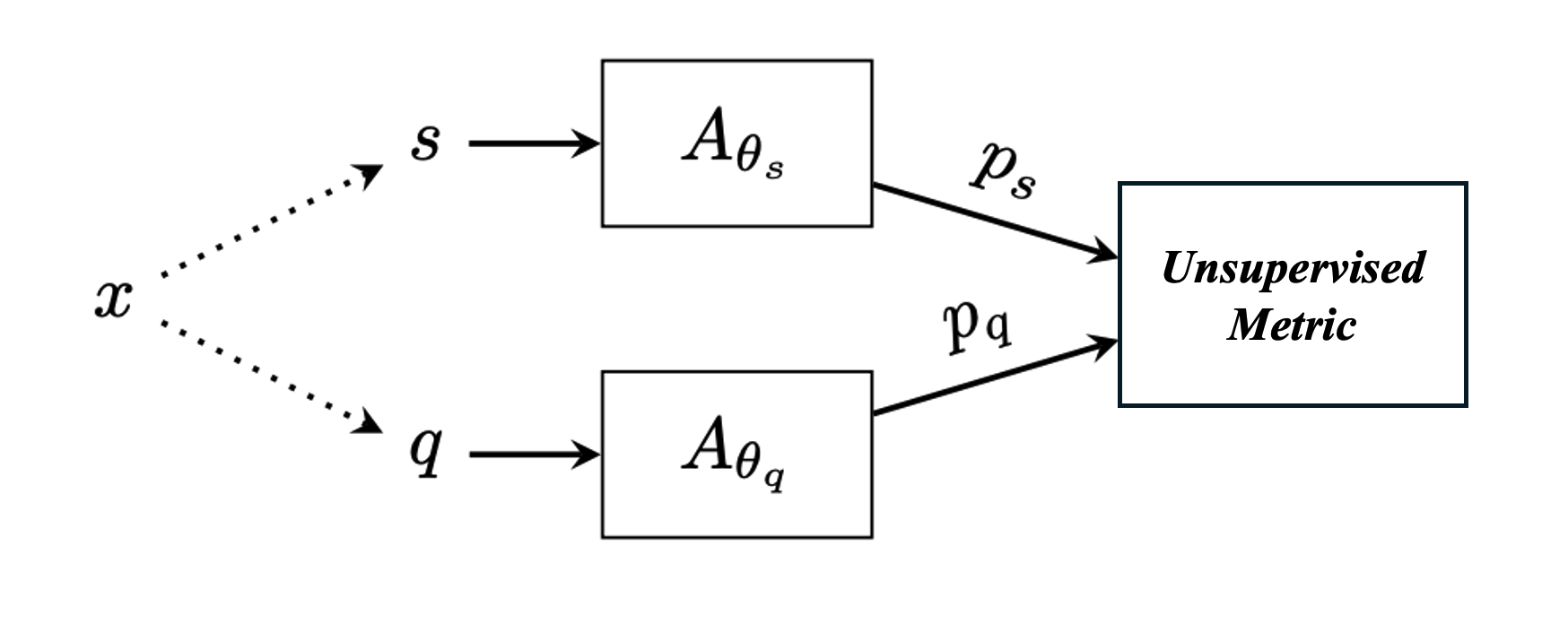}
    \caption{Generic CoAD schematic with an unknown state variable \( x \) influencing measured data inputs \( s \) and \( q \) fed to corresponding models, \( A_{\theta_s} \) and \( A_{\theta_q} \). The models are trained to optimize an unsupervised objective metric such as correlation, covariance, or $\hat{F}_{\beta} $ such that the models maximally separate \( x \) into normal and anomalous behavior~\cite{humble2024coincident}.
}
    \label{fig:coad_schematic}  
\end{figure}

From a technical perspective, coincident learning utilizes two models, \( A_{\theta_s}(s) \) and \( A_{\theta_q}(q) \), parameterized by \( \theta_s \) and \( \theta_q \), respectively. Each model outputs a scalar value, \( p_s \) and \( p_q \), where \( p_s, p_q \in [0, 1] \), representing the confidence that the corresponding input is anomalous. The marginal datasets are \( D_s = \{s\} \) and \( D_q = \{q\} \). Both \( s \) and \( q \) are assumed to be functions of an unknown variable \( x \), denoted as \( s(x) \) and \( q(x) \). A schematic representation of the method is shown in \cref{fig:coad_schematic}.

The goal of learning is to optimize \( A_{\theta_s} \) and \( A_{\theta_q} \) to maximize the alignment between the outputs \( p_s \) and \( p_q \). Previously proposed unsupervised metrics include covariance, correlation, and \( \hat{F}_{\beta} \).  Specifically, $\hat{F}_\beta $ is defined as follows: 
\begin{align}
\hat{F}_{\beta} =
(1 + \beta^2)
\;
\frac{
  \mu_{sq} - \mu_s \mu_q
}{
  \mu_{sq} + \alpha \beta^2
}
\;
\frac{
  1 - \mu_{sq}
}{
  (1 - \mu_s)(1 - \mu_q)
} \,,
\end{align}%
where $\alpha$ is the anomaly fraction in the data, $\mu_{sq} = E[A_{\theta_s}(s) A_{\theta_q}(q)]$, $\mu_s =  E[A_{\theta_s}(s)]$, and $\mu_q =  E[A_{\theta_q}(q)].$ An estimate of $\alpha$ is sufficient. It has been shown that $\hat{F}_\beta$ is a lower bound for ${F}_\beta$ in \cite{humble2024coincident}. In this work, we use the \( F_\beta \) objective with the choice of \( \beta = \infty \), which is equivalent to the covariance objective. This choice of \( \beta \) emphasizes recall and improves training stability.

\section{Experiments} 
\label{sec:results} 

\subsection{RF phase vs RF amplitude diagnostic data}

We evaluate the performance of our two-stage RF station anomaly detection method, shown in \cref{fig:two-stage-method}, using both RF phase and amplitude data over the period from January 19, 2024, to January 31, 2024. To simplify data processing, we focus exclusively on operational periods during which the beam was delivered to experiments under healthy conditions \cite{humble2022beam}. Additionally, we exclude time periods when the beam-synchronous acquisition (BSA) data-recording service exhibited anomalous performance, identified using the Bayesian Online Changepoint Detection (BOCPD) algorithm \cite{adams2007bayesian}. We follow the same data filtering protocol as \cite{humble2022beam}. After applying these criteria, we analyze approximately 230 hours of data within the specified period. 

Within the selected period, the proposed method generates 5,383 anomaly candidates using RF amplitude data and 5,364 candidates using RF phase data. After excluding system-level faults—faults in which all RF stations exhibit abnormal behavior, preventing attribution to specific stations— the proposed method using RF amplitude diagnostic data identifies 209 anomalies, 185 of which are confirmed by manual inspection by experts as true anomalies, resulting in a precision of 88.52\%. Similarly, the same method using RF phase diagnostic data predicts 548 anomalies, of which 482 are true anomalies, yielding a precision of 87.95\%. These results are presented in \Cref{tab:ampl_bpm_counts}.  Additionally, after examining 300 negative examples predicted by both methods, we observed 3 and 4 false negatives for RF amplitude data and RF phase data, respectively, implying that the false negative rate for each is approximately 1\%. Both methods predict approximately 5,000 examples as negatives, with around 50 events missed by each. This results in a recall \footnote{It is important to note that recall is calculated based on the anomaly candidate dataset generated in \cref{sec:anomaly_candidate_identification}.} of 87\% for the phase-based method and 78\% for the amplitude-based method. Next, we examined the overlap and distinctiveness of anomalies identified by both methods. We consider RF anomalies identified using amplitude diagnostic data to be equivalent to those identified using phase diagnostic data if the following conditions are satisfied: (1) the anomalies share the same time windows for the BPM data, and (2) the faults are associated with the same RF station. Based on these criteria, 123 anomalies were unique to RF amplitude diagnostic data, while 420 anomalies were unique to RF phase diagnostic data, with 62 anomalies identified by both RF amplitude and RF phase diagnostic data. Consequently, by integrating RF phase diagnostic data with amplitude diagnostic data, anomaly detection improves by more than threefold, identifying 605 anomalies using both data streams compared to just 185 detected using amplitude data alone.

\begin{figure*}[htbp]  
    \centering  
    \includegraphics[width=1\textwidth]{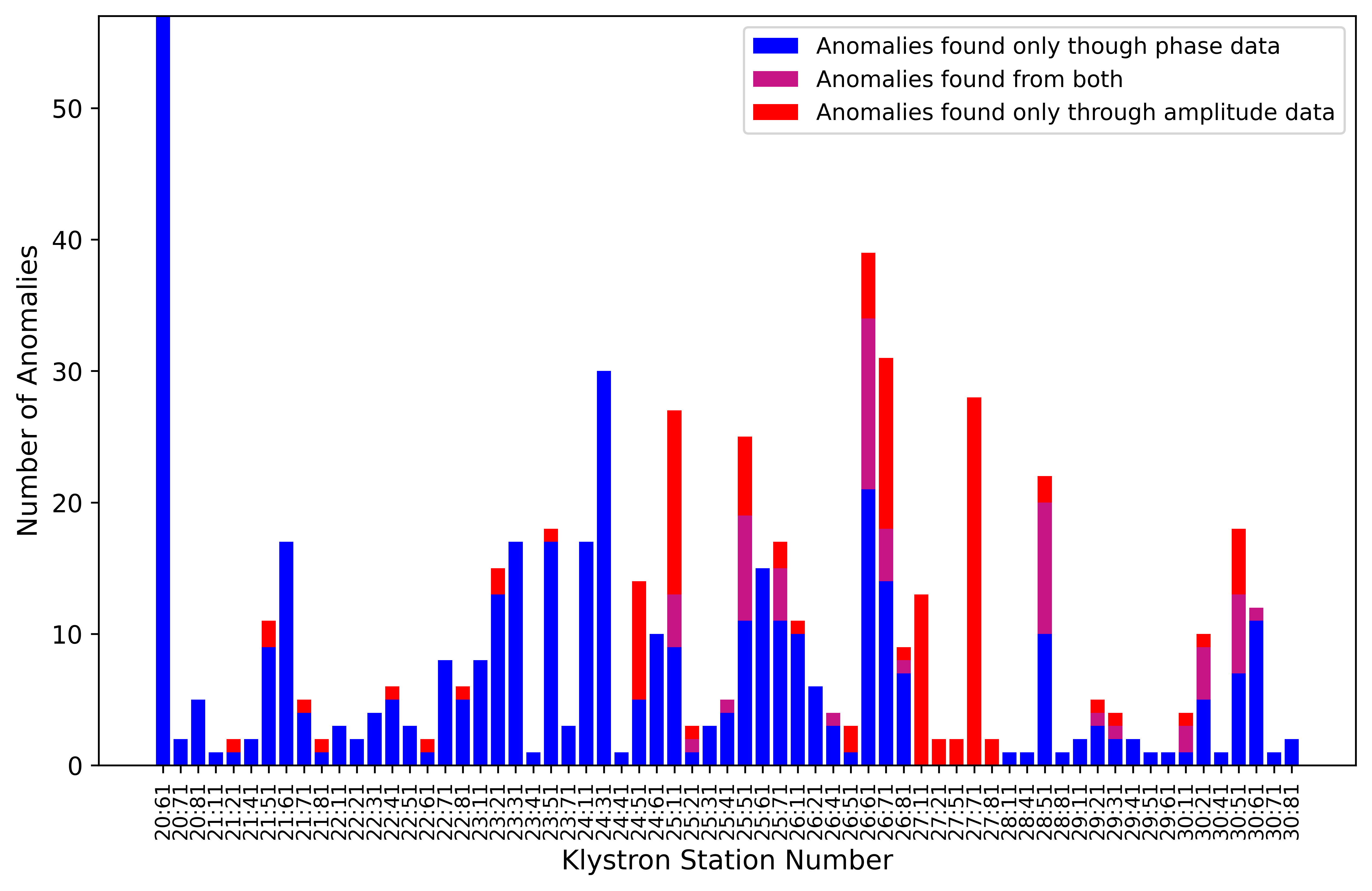}
    \caption{Distribution of the anomalies in RF stations found using CoAD. The colors correspond to the data source: red for anomalies found only from amplitude data, blue for anomalies found only from phase data, and magenta for anomalies found in both data streams. Sector 27 has no phase anomalies due to a software fault preventing logging of phase data during this time period. Note that the critical sector 20 injector station anomalies are only found from phase data.}
    \label{fig:ampl_vs_bpm}  
\end{figure*}

A detailed investigation of the distribution of anomalies across RF stations further underscores the benefits of incorporating RF phase diagnostic data. As illustrated in \cref{fig:ampl_vs_bpm}, anomalies identified exclusively by RF amplitude diagnostic data are predominantly concentrated in specific RF stations, such as LI24:51, LI25:11, LI26:71, and LI27:11--81. A closer analysis of LI27:11--81 reveals that phase data is absent across all LI27 RF stations, explaining why anomalies in these stations can only be detected using RF amplitude diagnostic data. In contrast, anomalies detected solely through RF phase diagnostic data are distributed across a broader range of RF stations. Notably, certain RF station faults, such as those in 20:61-81, can only be identified using RF phase data. These stations are particularly critical for anomaly detection, as they are located near the injector, where there are fewer spare RF stations to compensate for faulty behavior. This finding highlights that incorporating RF phase diagnostic data not only increases the overall detection capability but also enhances the diversity of detected anomalies in terms of RF station coverage.

An additional advantage of utilizing RF phase diagnostic data is its ability to identify some types of jitter, e.g., the appearance of frequent, small spikes in phase, which would be undetectable using amplitude data alone. Further details on repeated spike faults are provided in \cref{sec:jitter_detection}. 

\begin{table}[htbp]
\centering
\begin{tabular}{@{}lrrr@{}}
\toprule
                          & \# of Anomalies & Precision  & Recall \\ \midrule
Phase-based               & 482             & 87.95\%  & 87\% \\
Amplitude-based                & 185             & 88.52\%  & 78\% \\ \bottomrule
\end{tabular}
\caption{Performance evaluation of CoAD in detecting RF anomalies using phase vs.\ amplitude diagnostics data: A case study from January 19 to January 31, 2024. Note that 62 anomalies can be identified by both RF amplitude and RF phase diagnostic data and are accounted for in both lines. }
\label{tab:ampl_bpm_counts}
\end{table}

\subsection{Comparison of different anomaly detection methods}

For both the RF amplitude diagnostic data and RF phase diagnostic data, we employ CoAD as our de facto anomaly detection method. In this section, we evaluate the effectiveness of CoAD in comparison to other anomaly detection methods. Specifically, we focus on RF phase data for this evaluation. The following list details the methods that are compared: 

\begin{itemize}
    \item CCA-1D: Canonical Correlation Analysis (CCA) is a statistical method employed to identify and quantify the correlation between two sets of variables. We utilize the implementation provided by the library \cite{bilenko2016pyrcca}. To apply this method, it is necessary to convert our 2D data into a 1D format. We include the CCA method because it can be regarded as the linear counterpart of CoAD.
    \item CCA-2D: Similar to the above, except this method accommodates 2D data. We employ the implementation from the library \cite{min2019tensor}.  
    \item IForest \cite{liu2008isolation}: an unsupervised machine learning algorithm that identifies anomalies by recursively partitioning the data using randomly selected features and split values, effectively isolating outliers due to their shorter paths in the tree structure. 
    \item OCSVM \cite{manevitz2001one}: an unsupervised learning algorithm that constructs a decision function for outlier detection by estimating a high-dimensional boundary around the data's normal class, effectively separating outliers based on their deviation from this boundary.
    \item DGHL \cite{challu2022deep}: employs a top-down convolutional network to map multivariate time-series windows to a hierarchical latent space, using alternating backpropagation and short-run MCMC for training by maximizing observed likelihood, thus enabling efficient anomaly detection and robust handling of missing data.
    \item OmniAnomaly \cite{su2019robust}: a robust multivariate time series anomaly detection method using a stochastic recurrent neural network that leverages techniques like stochastic variable connection and planar normalizing flow to learn and reconstruct normal patterns.
\end{itemize}

AUCPR and best $F_1$ score \footnote{It is important to note that both AUCPR and the best $F_1$ score are calculated based on the anomaly candidate dataset generated in \cref{sec:anomaly_candidate_identification}.} are commonly used metrics to evaluate the performance of anomaly detection algorithms. In \Cref{tab:perform_aucpr_bestf1}, we report both metrics for all previously discussed anomaly detection methods. These metrics do not require selecting a decision threshold—they summarize performance across all possible thresholds. However, in deployment scenarios, a specific threshold must be chosen to classify inputs as normal or anomalous. In Table 4, we examine how sensitive each method is to the choice of threshold and show that CoAD maintains stable performance across a wide range of thresholds. All experiments use 70\% of the data for training and 30\% for testing.

Our data consists of two input streams: RF diagnostic data and BPM data. Methods like CCA-1D, CCA-2D, and CoAD naturally accommodate both input streams. However, methods such as OmniAnomaly, DGHL, IForest, and OCSVM are designed to process only a single stream of data by default. To benchmark against methods not designed for multiple inputs, we attempted two strategies: Score \& Stack (training and scoring on each input independently and using the Pareto frontier of all threshold pairs) and Stack \& Score (training and scoring on the stacked input) \cite{humble2024coincident}. For these methods, we ran both strategies and reported the score of the better-performing approach. The performance of all the methods using AUCPR and best $F_1$ metrics is shown in \Cref{tab:perform_aucpr_bestf1}. The table indicates that OCSVM achieves the highest performance among all traditional methods. OmniAnomaly outperforms all traditional methods except OCSVM, while DGHL ranks as the second-best method overall. Notably, CoAD demonstrates the best performance across all methods, achieving the highest AUCPR and best $F_1$ scores.

In real-world applications, setting a threshold is essential to convert anomaly scores into anomaly predictions. Therefore, a fair evaluation also requires the calculation of the $F_1$ score using various thresholding methods. To evaluate the performance of different anomaly detection methods using $F_1$ score under various thresholding strategies, we employ the following thresholding methods: Filter \cite{hashemi2019filtering}, Meta \cite{zhao2020automating}, and MixMod \cite{veluw2023application}. All of these thresholding methods are implemented in the Pythresh library, and the library's comprehensive benchmark across various datasets and methods demonstrates that these chosen thresholding methods achieve superior performance. Since the $F_1$ score with thresholding methods will be worse than the best $F_1$ score, we calculated the $F_1$ score with various thresholding methods only for the deep learning methods. The results are shown in \Cref{tab:peform_f1_threshold}. From the table, it is evident that the CoAD method achieves good results with all three thresholding methods. Moreover, its F1 scores are only slightly lower than the best $F_1$ score. In contrast, the $F_1$ scores for OmniAnomaly and DGHL with various thresholding methods are significantly worse than their best $F_1$ scores. This result indicates that CoAD is more robust at separating the anomalous and normal samples, thereby increasing the probability of success during deployment. Furthermore, because CoAD provides an unsupervised estimate of precision and recall, the CoAD framework can also be used to select a threshold for deployment even in the absence of labels.

\begin{table}[htbp]
\centering 
\begin{tabular}{@{}llrr@{}}
\toprule
                                   &               & AUCPR & Best $F_1$ \\ 
                                   \midrule
\multicolumn{1}{l}{Traditional}   & CCA-1D        & 0.38                      & 0.46                         \\
\multicolumn{1}{l}{}              & CCA-2D        & 0.44                      & 0.49                         \\
\multicolumn{1}{l}{}              & IForest       & 0.42                      & 0.43                         \\
\multicolumn{1}{l}{}              & OCSVM         & 0.63                      & 0.65                         \\ \midrule
\multicolumn{1}{l}{Deep Learning} & DGHL          & 0.82                      & 0.76                          \\
\multicolumn{1}{l}{}              & OmniAnomaly & 0.49                      & 0.53                         \\
\multicolumn{1}{l}{}              & CoAD          & \textbf{0.93}                       & \textbf{0.85}                         \\ 
\bottomrule
\end{tabular}

\caption{Comparison of different anomaly detection methods on phase-based diagnostic data. CoAD shows superior performance under both AUCPR and $F_1$.}
\label{tab:perform_aucpr_bestf1}

\end{table}

\begin{table}[htbp]
\centering
\begin{tabular}{@{}lrrr@{}}
\toprule
                             & Meta & Mixmod & Filter \\ \midrule 
CoAD                         & \textbf{0.84}                   & \textbf{0.82}                    & \textbf{0.81}                     \\
OmniAnomaly  & 0.5                      & 0.5                        & 0.5                      \\
DGHL           & 0.36                    & 0.63                      & 0.3                      \\ \bottomrule 
\end{tabular}
\caption{Comparison of all the deep learning based methods using $F_1$ score with three different thresholding methods: Meta, Mixmod, and Filter. In all cases, CoAD produces the best scores.}
\label{tab:peform_f1_threshold}
\end{table}

\subsection{Clustering and root-cause analysis}

A key advantage of RF phase data is the high information content of the synchronous, beam-rate data compared to the slow, asynchronous amplitude diagnostic. The complexity of the RF phase data could reveal signatures of the underlying root cause, i.e., the ability not just to identify a specific problematic RF station, but also \textit{why} that station failed. Providing this information to operators would improve both the interpretability of warnings and the speed of diagnosis and recovery from faults.

\begin{figure}[htbp]  
    \centering  
    \includegraphics[width=0.5\textwidth]{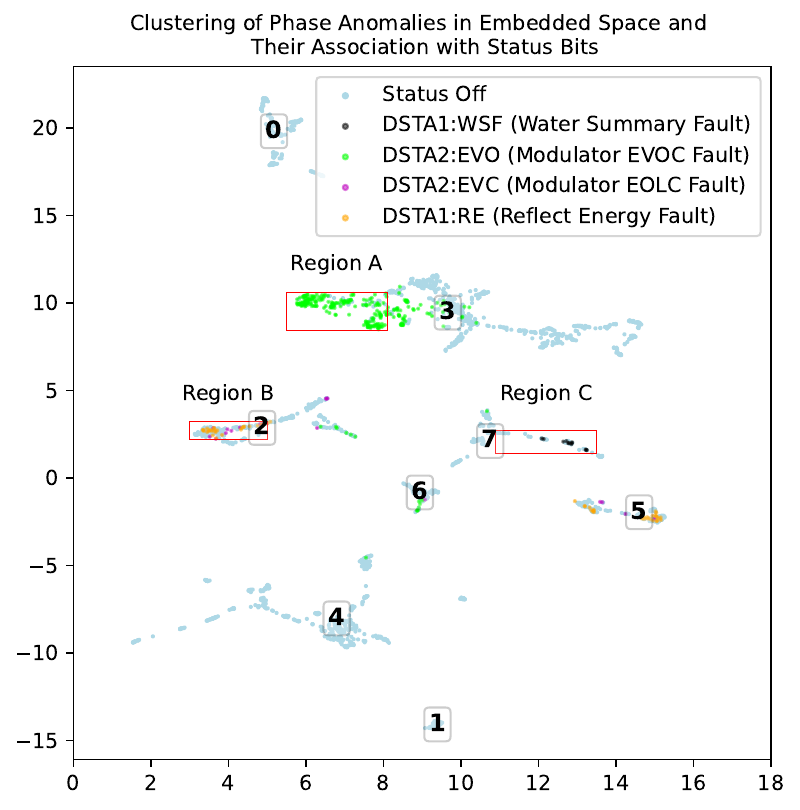}
    \caption{Clustering of the anomalous RF phase signals and their association with status bits. Example phase and BPM data selected from regions A, B, C are shown in \cref{fig:detailed look of the phas_fast signals}, with more details shown in \cref{fig:zoomed_view_cluster_status}.}
    \label{fig:clustering_status}  
\end{figure}

To investigate the potential for root-cause analysis from RF phase diagnostic data, we undertake the following steps: We begin by collecting the RF phase signals labeled by CoAD as anomalous from the following time periods: February 7–23, 2022; March 5–17, 2022; April 1–16, 2022; November 9–18, 2023; November 21–December 1, 2023; and January 19–31, 2024. We then apply UMAP \cite{mcinnes2018umap} to reduce the dimensionality of the raw signals to two dimensions and employ HDBSCAN \cite{mcinnes2017hdbscan} to cluster the 2D embeddings. The clusters are displayed in \cref{fig:clustering_status}. 

More importantly, in the same figure, we also show the relationship between the clusters and their association with different `status bits.' The monitoring system for each RF station records not only amplitude and phase information but also various status bits, including Digital Status 1 (DSTA1), Digital Status 2 (DSTA2), Status (STAT), Hardware Descriptor (HDSC), Hardware Status (HSTA), and PIOP Status (SWRD). Each status bit further contains sub-status indicators, which are intended to reveal the underlying cause of a hardware problem. For example, \texttt{DSTA1:RE} signifies the Digital Status Reflected Energy fault, while \texttt{DSTA2:EVO} denotes a Modulator EVOC Fault. While highly specific, these status bits are known to have poor precision and recall, and thus cannot be used as an anomaly detection system by themselves.

To better understand fault patterns, we analyzed the activation and deactivation of these status bits during RF faults. While many status bits exhibited interesting trends, our discussion primarily focuses on the following: \texttt{DSTA1:RE}, \texttt{DSTA1:WSF}, \texttt{DSTA2:EVC}, and \texttt{DSTA2:EVO}. The distribution of these status bits across clusters is also illustrated in \cref{fig:clustering_status}. The same figure also highlights several regions with red boxes, and their zoomed views are shown in \cref{fig:zoomed_view_cluster_status} in \cref{sec:zoomed_view}. 

\begin{figure*}[htbp]
    \centering
    \subfigure{
        \includegraphics[width=1\textwidth]{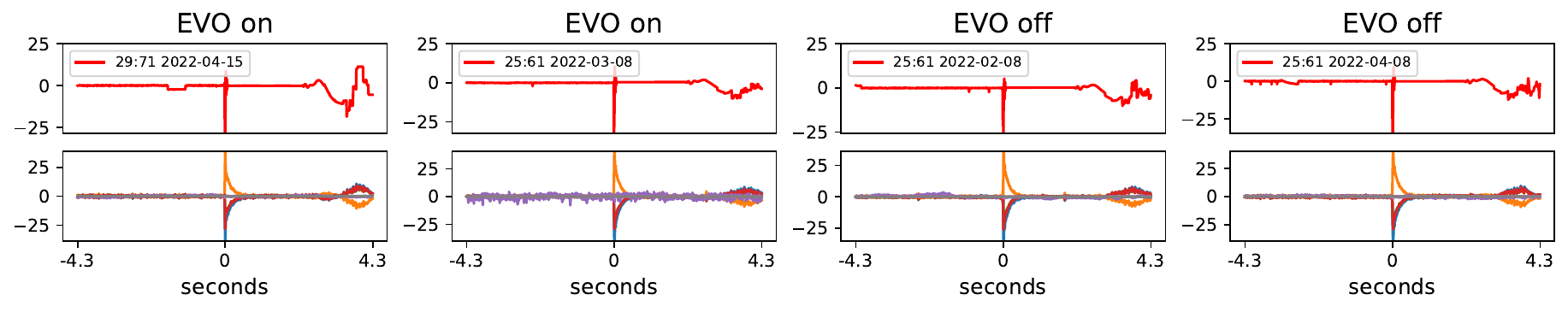}
        \label{fig:graph1}
    }
    
    (a)

    \subfigure{
        \includegraphics[width=1\textwidth]{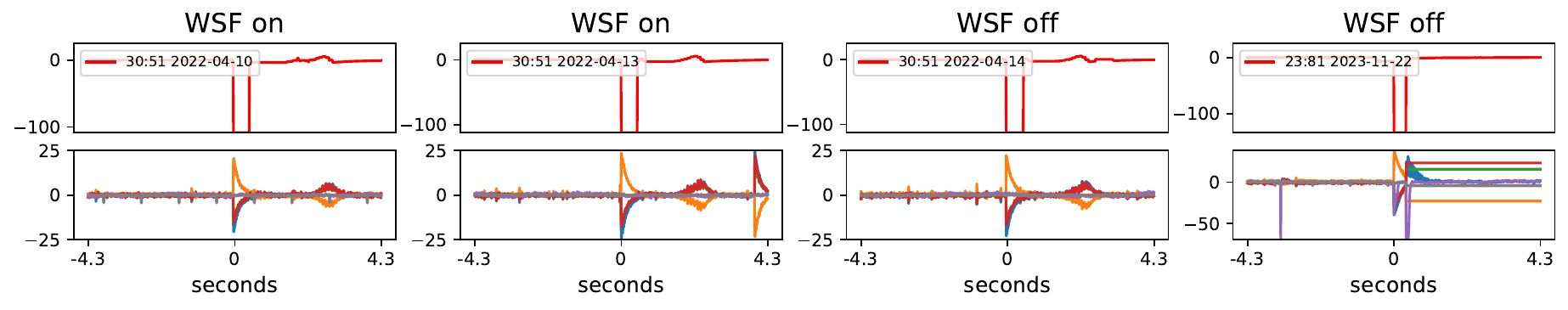}
        \label{fig:graph2}
    }
    
    (b)

    \subfigure{
        \includegraphics[width=1\textwidth]{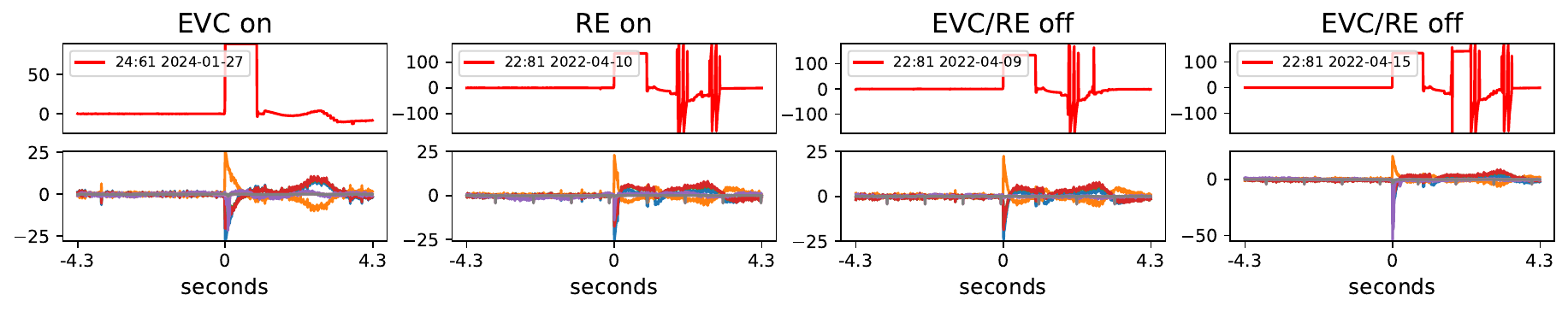}
        \label{fig:graph3}
    }
    
    (c)

    \caption{Details of the phase signals associated with different status bits in \cref{fig:clustering_status}. Four examples are shown for each category, with each example consisting of two components: the anomalous phase signals (shown in the first row) and the corresponding BPM signals (displayed in the second row). EVC and RE status bits are shown together due to similarity of behavior.}
    \label{fig:detailed look of the phas_fast signals}
    
\end{figure*}

\begin{itemize}

    \item \textbf{\texttt{DSTA2:EVO} (Modulator EVOC Fault---Electrode Voltage Overcurrent)}: 
    The klystron is powered by a high-voltage pulse supplied by the modulator. This pulse is increased to the needed level by a transformer. The voltage and current of this pulse are monitored by sensors and reported to the control system. If the current in the klystron pulse is too high, a safety system called the EVO interlock will shut down the modulator to prevent damage. In older modulators, a series of high-current signals can indicate a problem inside the klystron’s pulse tank. This could mean that the modulator is producing a higher voltage than it should. By checking these signals on an oscilloscope (a device that shows electrical signals over time), you can see exactly when and where the issue occurs. A sudden spike in the signal could be caused by an electrical arc (a high-voltage discharge) somewhere in the high-voltage system, which can also trigger a fault.

   This fault predominantly appears in the left part of Cluster 3. Notably, part (a) of \cref{fig:detailed look of the phas_fast signals} reveals a distinct signature of RF phase associated with EVO bits---a wave-like motion occurring 3 to 4 seconds after the initial spike at the window’s center. This characteristic oscillation may serve as a unique indicator of the EVO fault.  

    \item \textbf{\texttt{DSTA1:WSF} (Water Summary Fault)} happens when any of the various flow switches that monitor cooling water to the station are faulted. 

    This fault appears in the right region of Cluster 7 and is characterized by a sharp, rectangular drop of more than -100 degrees lasting approximately 0.5 seconds. See part (b) of \cref{fig:detailed look of the phas_fast signals} for more details.

    \item \textbf{\texttt{DSTA2:EVC} (Modulator EOLC Fault---End of Line Clipper)}:
     happen when too much current flows due to high reverse voltage in the system. A small amount of reverse voltage helps the thyratron switch off properly, but if it exceeds 5–6 kV, it can damage components. A protective circuit (thyrite stack) limits this voltage but increases current, which can trigger the EOLC fault and shut down the modulator to prevent damage. 
    
    EOLC faults can happen when there is a mismatch between the PFN (Pulse Forming Network) and the klystron setup. Poor tuning or failing PFN capacitors can cause this, which can be seen on a time plot. Arcing in the pulse cable, usually near the PFN or charge inductor, can also trigger faults — and it’s often loud enough to hear. Less common issues include problems with the core bias reset circuit, which can cause large voltage reversals, or in rare cases, failures in the pulse transformer or klystron. However, if the klystron is working reliably most of the time, it’s unlikely to be a major issue.

    These faults appear in parts of Cluster 2 and Cluster 5. In the phase signal, they show up as a sharp rectangular jump starting in the middle of the time window and lasting about 1 second, with a magnitude of about 100 degrees or more. After the jump, a wave-like motion follows for about 2 seconds. Refer to part (c) of \cref{fig:detailed look of the phas_fast signals} for more details.

    \item \textbf{\texttt{DSTA1:RE} (Klystron Reflected Energy Fault)}:  
     happens when too much energy is reflected back into the klystron instead of being transferred to the load (like a particle accelerator or radar system). Klystrons are designed to amplify and transmit high-power microwave signals, but if the signal is not absorbed properly by the load, the energy can reflect back into the klystron and cause damage or instability. Common causes include load mismatches, waveguide arcing, vacuum leaks, power overload, and component misalignment or failure. Protective systems detect these faults and shut down the modulator to prevent damage.

    As seen in parts of Cluster 2 and Cluster 5, these faults show up in similar regions to those of EVC Fault. Part (c) of \cref{fig:detailed look of the phas_fast signals} shows a sharp, rectangular jump starting midway through the time window, lasting about 1 second with a magnitude of around 100 degrees or more. This is followed by a wave-like motion lasting about 2 seconds. It is common to see EVC and RE faults together, possibly caused by ringing in the klystron or pulse tank that feeds back into the PFN in the modulator.

\end{itemize}

To further investigate the status bits, we focused on three distinct regions, denoted as Region A, Region B, and Region C, as shown in \cref{fig:clustering_status}. A magnified view of these regions is provided in \cref{fig:zoomed_view_cluster_status}.

In Region A, we observed that 85\% of the anomalous samples had the DSTA2:EVO status bit activated. This finding reinforces a strong correlation between the physical signature in the phase fast signal within this region—specifically, a wave-like motion occurring 3 to 4 seconds after the initial spike at the center of the window—and the Modulator EVOC Fault, which corresponds to an Electrode Voltage Overcurrent.

In Region B, we found that 8\% and 2\% of phase anomalies were associated with the DSTA2:EVC and DSTA1:RE status bits, respectively. While these percentages are relatively low, it is possible that the low detection rate reflects the low recall of these status bits. Although it is challenging to definitively conclude that the inactive status bits represent missed cases, we found that approximately 40\% and 9\% of the anomalies are associated with the DSTA2:EVC and DSTA1:RE status bits, respectively, under the following circumstances: the anomalies occurred in the same klystron and within a 24-hour window of another anomaly where the status bit was active. Thus, anomalies within parts of Cluster 2 and Cluster 5 may be indicative of DSTA2:EVC or DSTA1:RE faults.

In Region C, 16\% of anomalies had the DSTA1:WSF status bit activated. Applying the same criteria as in the previous case, we found that approximately 60\% of anomalies in this region occurred in the same klystron and within the same 24-hour window as another anomaly where the WSF status bit was active. Notably, the majority of anomalies in this region were traced to station LI30:51. Additionally, data from the CATER system \cite{rogind2012integrated} at SLAC—a system where operators manually record faults and their resolutions—indicated that station 23:81 experienced a water leak fault, and the embeddings of its anomalous phase signals lie within Region C.

As shown in \cref{fig:clustering_status}, nearly all Modulator EVOC Faults are grouped in Cluster 3, while Modulator EOLC Faults and Reflected Energy Faults are primarily associated with Clusters 2 and 5. Similarly, Water Summary Faults are almost exclusively located in Cluster 7. These associations suggest that the clustering results are indicative of underlying fault mechanisms. Moreover, each cluster exhibits distinct physical signatures in the phase data, further implying that phase patterns are reflective of the corresponding root causes. Notably, in the regions linked to the aforementioned clusters (Regions A, B, and C), the relevant status bits are not always active. However, as illustrated in \cref{fig:detailed look of the phas_fast signals}, samples with inactive status bits still display anomalous behaviors and phase signatures similar to those observed when the status bits are active. This observation suggests that status bits have limited recall, likely due to their lower sampling rate (0.2 Hz) relative to the phase data (120 Hz). Overall, phase data enables the detection of substantially more fault instances and provides physical signatures that are potentially diagnostic of root causes. Even when clusters contain faults associated with multiple status bits and distinct underlying mechanisms, they offer valuable operational insight by narrowing the scope of possible causes. The presence of active status bits in many cases further supports that phase-based anomalies correspond to genuine hardware issues.

\subsection{Interpreting CoAD model with shapley value}


During the anomaly candidate identification step (\cref{sec:anomaly_candidate_identification}), the onset of the anomaly was positioned at the center of the time window. As a result, anomalies in the RF phase signal and BPM are expected to manifest at or near the center of the time window. Since most anomaly detection models provide only a single label for the entire time window, the precise timing of anomalies is not specified in the output.

\begin{figure}[htbp]
    \centering
    \subfigure{
        \includegraphics[width=0.5\textwidth]{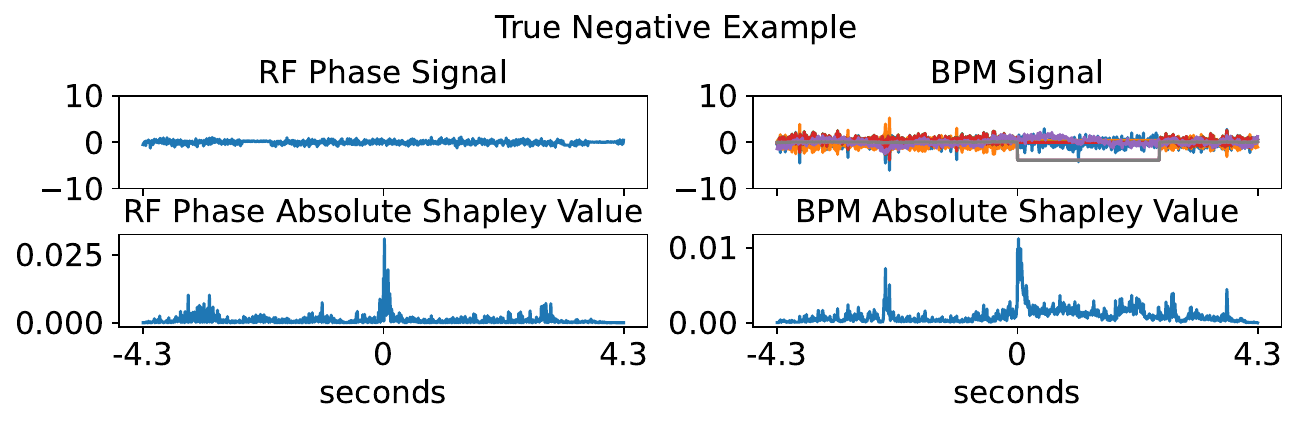}
    }
    
    (a)
    \subfigure{
        \includegraphics[width=0.5\textwidth]{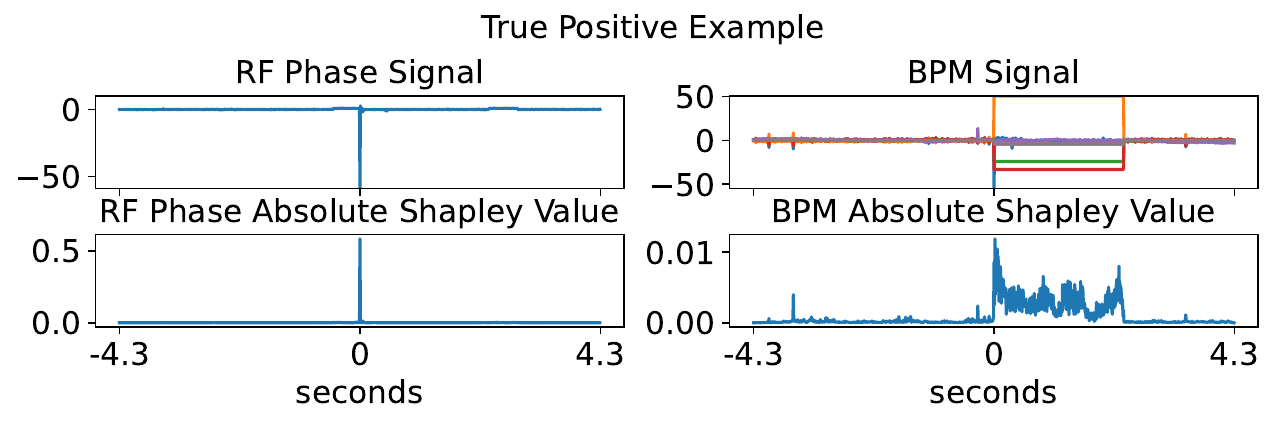}
    }
    
    (b)
    \subfigure{
        \includegraphics[width=0.5\textwidth]{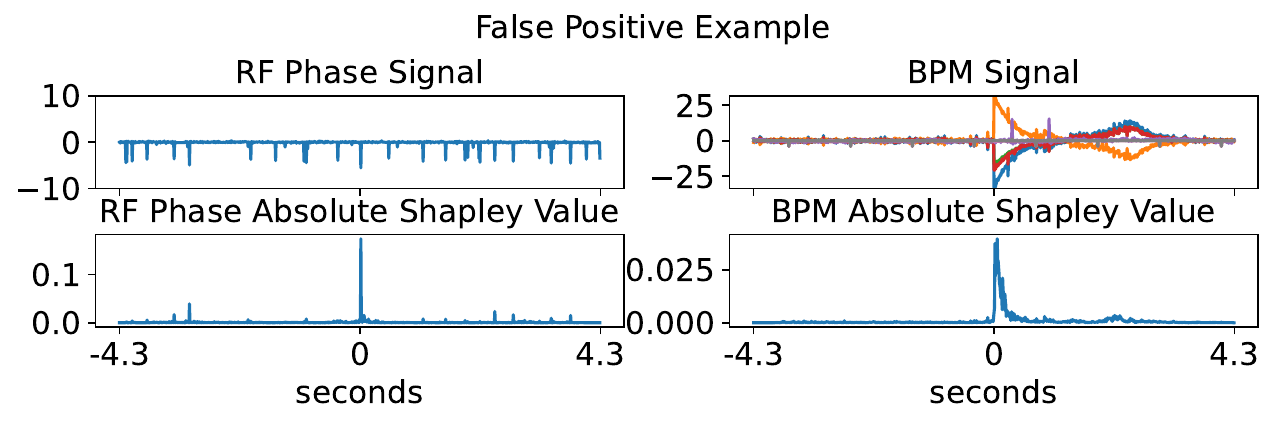}
    }
    
    (c)
    \caption{Shapley value-based explanation of CoAD predictions. Subplots (a) and (b) show examples of correct predictions, with (c) showing a false positive. The model primarily focuses on the center of the time window, consistent with the candidate generation process which attempts to place fault initiation at $t = 0$. In subplot (c) the model disregards broader statistical properties, such as the standard deviation of the time window, resulting in a false positive.}
    \label{fig:shap_value}
    
\end{figure}

Interpretable machine learning methods are designed to elucidate how models make predictions. In this context, an effective model should exhibit a strong focus on the center of each time window. This central focus should be evident in the results of interpretable methods. Therefore, to assess whether our CoAD algorithms are correctly attending to the relevant temporal regions during prediction, we employ Shapley values \cite{lundberg2017unified} to analyze model behavior. The Shapley value, a concept from cooperative game theory, allocates a fair share of the total contribution to each participant based on their individual inputs. For each pair of RF phase and BPM signals, we use the SHAP library \cite{lundberg2017unified} to compute Shapley values during the inference process, assigning a single Shapley value to each timestamp. Subsequently, we compute the absolute values of the Shapley scores. Examples of a normal and an abnormal case are illustrated in part (a) and (b) of \cref{fig:shap_value}, respectively. These figures demonstrate that the model predominantly attends to the center of the time window for anomaly predictions, which aligns with expected behavior. This observation indicates that the CoAD method performs effectively in focusing on the correct temporal regions for anomaly detection.

A more insightful application of Shapley values is to understand why the model makes incorrect predictions. As illustrated in part (c) of \cref{fig:shap_value}, there is an example where the model incorrectly classifies a normal sample as an anomaly. The sample is, in fact, normal because the RF phase spike at the center of the time window is part of the background noise in the RF phase, rather than indicative of anomalous behavior. This is evident from the presence of multiple spikes within the time window, suggesting that the RF station was experiencing consistent spikes during that period. Consequently, the spike at the center of the window is not due to a genuine anomaly in the RF phase. However, Shapley value analysis reveals that the model primarily focuses on the center of the window when making predictions, while disregarding the spikes in other parts of the time window. This indicates that the model overlooks broader statistical properties, such as the standard deviation of the time window, and instead singularly concentrates on the central region. This insight highlights a potential improvement for CoAD, suggesting that incorporating the standard deviation of the time window in the RF phase could enhance the model's performance in distinguishing true anomalies from background noise.

\section{Conclusion} 
\label{sec:conclusion} 

The inherent complexity of RF phase data has historically hindered its application to anomaly detection at SLAC. While classical methods fail to extract value from phase data, in this work we use the CoAD framework to train neural networks that operate on phase data. We show that deep learning models operating on phase data detect almost three times as many anomalies compared to models using amplitude data alone. We also find that CoAD outperforms existing traditional and deep learning methods. Moreover, the additional anomalies detected through phase data significantly increase the coverage of RF stations compared to amplitude-based detection, particularly in the critical injector region.

The rich information in phase data suggests that in addition to indicating the occurrence of anomalies, it can also provide more granularity in the prediction of the root cause of failure. We apply dimensionality reduction and clustering methods to identify distinct groups within the anomalous events identified by CoAD.  By integrating data from SLAC’s CATER system and status bits, we show that some individual clusters could correlate with specific RF station failure mechanisms, providing additional information to operators to support recovery from failures.

We also investigated the interpretability of the networks using Shapley values. Through these values, we observed that each neural network within the CoAD framework focused on the center of its respective time window, aligning with expectations based on the data generation process. These findings affirm that our CoAD model is functioning as intended and also provide insight into situations in which CoAD makes mistakes, informing future avenues of research. In future work, we plan to investigate the detection of repeated spike faults in the phase data, which currently remain undetectable by CoAD. Additionally, while CoAD presently operates on two data streams, we aim to extend its framework to accommodate multiple streams by leveraging graph neural networks. More broadly, the successful application of CoAD to this complex system highlights its potential for scalable anomaly detection across the thousands of subsystems within linear accelerators.

\begin{acknowledgements}
We thank Jim Craft, Xupeng Chen, Howard V. Smith, Benjamin Ripman and Timothy Maxwell for their helpful discussions. Use of the LINAC Coherent Light Source (LCLS), SLAC National Accelerator Laboratory, is supported by the U.S. Department of Energy, Office of Science, Office of Basic Energy Sciences under Contract No. DE-AC02-76SF00515.
\end{acknowledgements} 

\appendix 

\section{REPEATED SPIKE FAULT}
\label{sec:jitter_detection} 

\begin{figure}[htbp]  
    \centering  
    \includegraphics[width=.5\textwidth]{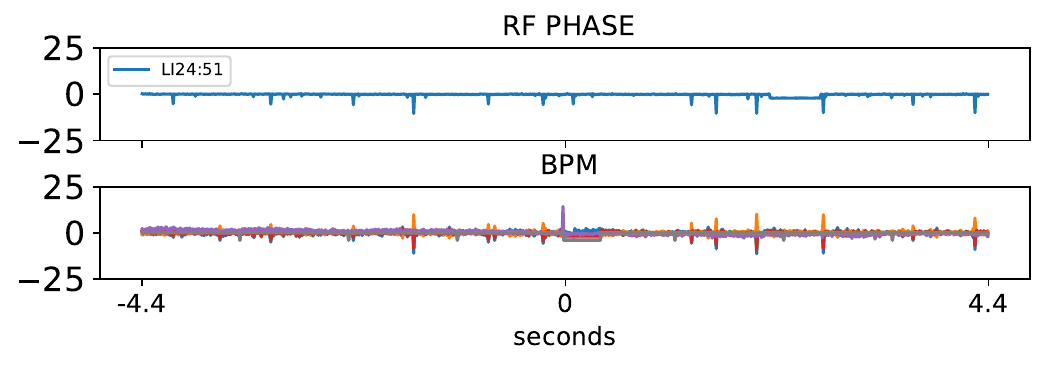}
    \caption{An example of repeated spikes fault.}
    \label{fig:mpnc}  
\end{figure}

Using synchronous RF phase diagnostic data, we identified a type of event that would be overlooked when only considering RF amplitude data. We refer to this special event as the repeated spike fault. These events are visualized in \cref{fig:mpnc}. As shown, these faults are sufficiently small that they do not exceed the BPM candidate-selection threshold, meaning that they are not considered candidates for fault detection in the first step of our method, shown in \cref{fig:two-stage-method}. This is also why those repeated spikes do not appear at the center of each time window. 

However, despite their small magnitude, repeated spike faults can occur at high frequency, still degrade beam quality, and are therefore undesirable. Consultation with domain experts revealed that these faults may result from thyratron ranging issues. In practice, such spike faults are generally considered tolerable provided that each spike occurs at intervals of no less than 23 seconds. As illustrated, these faults can repeat multiple times within each 8-second window, occurring simultaneously in both the BPM signals and the RF phase signals. Our investigation revealed that these faults primarily originated from RF stations LI:24:51 and LI:22:71 during the period from January 19, 2024 to January 31, 2024. Specifically, most of the faults from LI:24:51 are concentrated on 2024-01-23, while the majority of faults from LI:22:71 are concentrated on 2024-01-29. We estimate that the repeated spike fault occurred more than 10,000 times in LI:24:51 on 2024-01-23, and more than 2,000 times in LI:22:71 on 2024-01-29. 

The current candidate generation procedure described in \cref{sec:anomaly_candidate_identification} excludes repeated spike faults, as these faults typically do not surpass the threshold required for candidate selection. Consequently, these repeated spike faults often do not appear at the center of the data window and are labeled as normal samples during the labeling process.
Since our primary focus is on identifying hard faults rather than persistent instability, we refrain from adjusting the threshold to capture these events. Reducing the threshold to account for persistent instability would significantly increase the number of candidate samples, as repeated spike faults can occur thousands of times within a single day. 

\section{ZOOMED VIEW OF FIGURE 5}

\label{sec:zoomed_view}

In the following figure \cref{fig:zoomed_view_cluster_status}, we present a zoomed-in view of Regions A, B, and C from \cref{fig:clustering_status}, each corresponding to distinct fault types: Region A captures Modulator EVOC Faults (EVO), Region B includes both Modulator EOLC Faults (EVC) and Reflected Energy Faults (RE), and Region C corresponds to Water Summary Faults (WSF). Within each region, we highlight several representative examples with status bits both activated (on) and deactivated (off).

A key observation across all regions is that anomalous phase signatures appear similar regardless of the status bit state. For example, in Region A, both on- and off-bit cases exhibit a distinctive wave-like motion in the phase fast signal, occurring approximately 3–4 seconds after a sharp spike centered in the window. Additionally, these on-bit and off-bit cases are located near each other in the UMAP embeddings. This spatial proximity, combined with the similarity of their phase signatures, suggests that the off-bit cases likely represent true anomalies that were missed by the status bits but identified through the phase-based analysis. This, in turn, implies that the status bits have low recall. 
Overall, these findings indicate that phase-based anomaly detection identifies genuine faults that are sometimes missed by status bits. The consistency of the physical signatures—regardless of status bit activation—and their alignment with known fault types indicate that these anomalies correspond to real underlying issues in the RF system.

\begin{figure*}[htbp]

    \centering
    \subfigure{
        \includegraphics[width=0.99\textwidth]{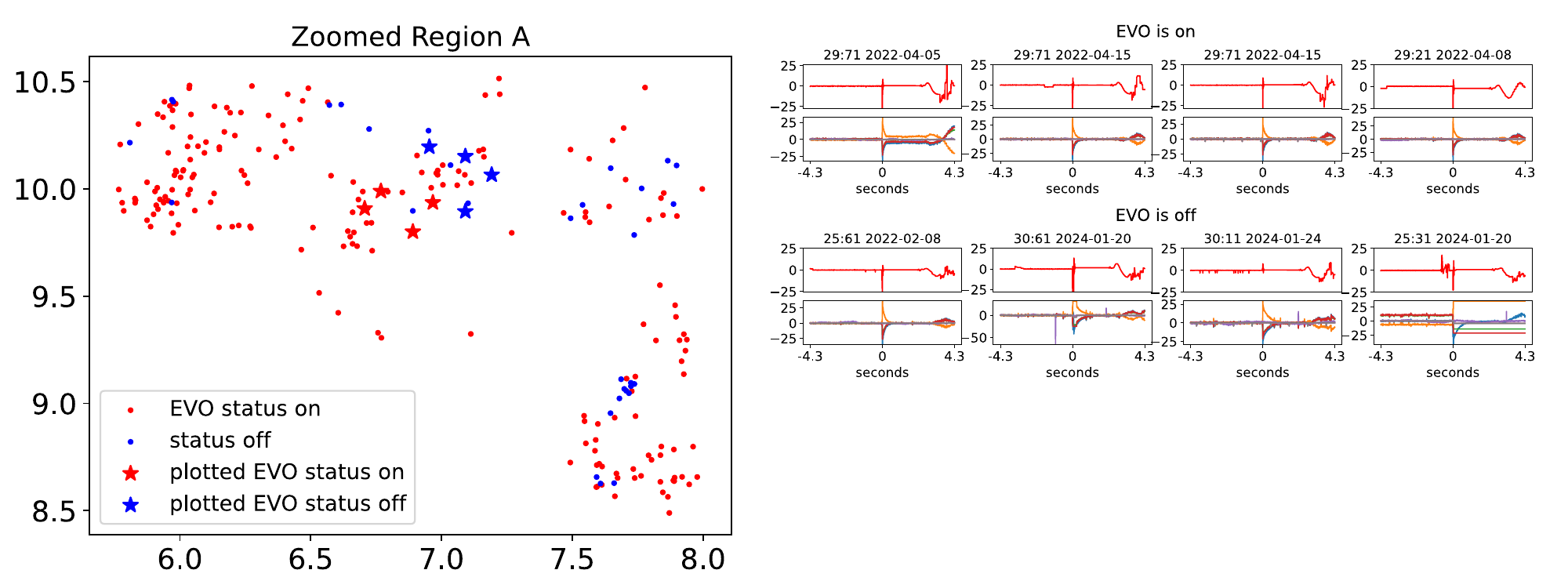}
    }
    \subfigure{
        \includegraphics[width=0.99\textwidth]{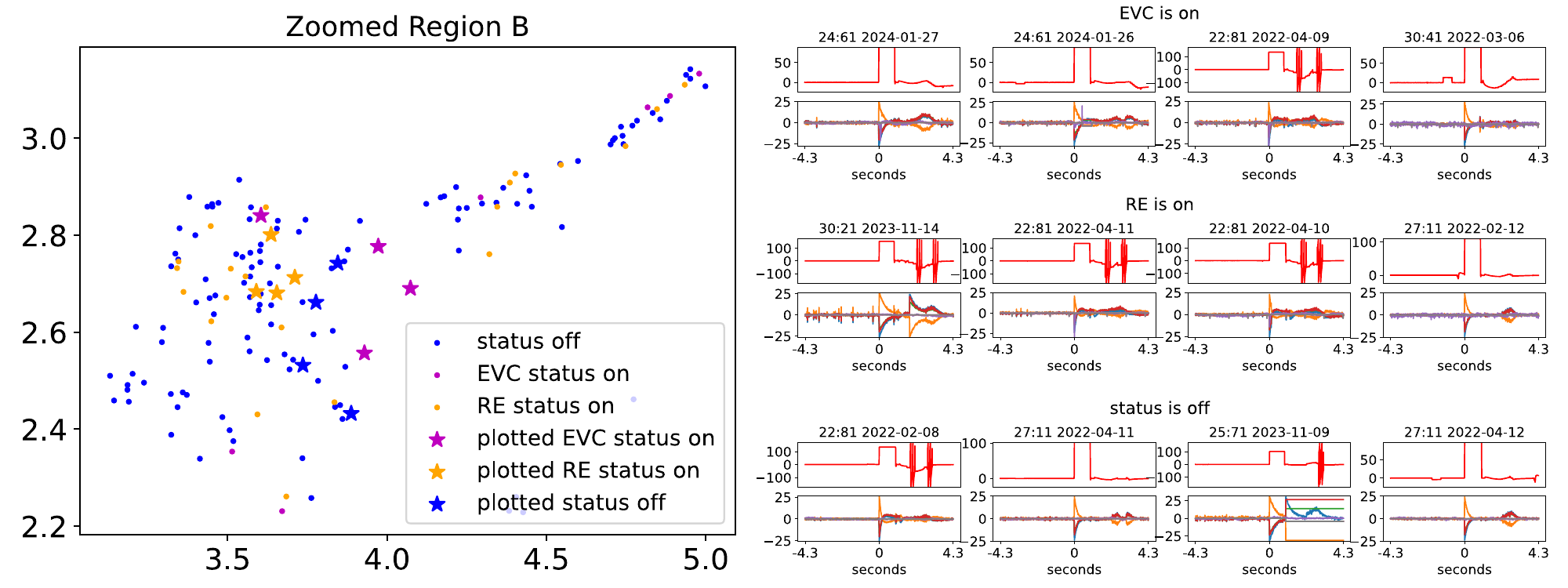}
    }
    \subfigure{
        \includegraphics[width=0.99\textwidth]{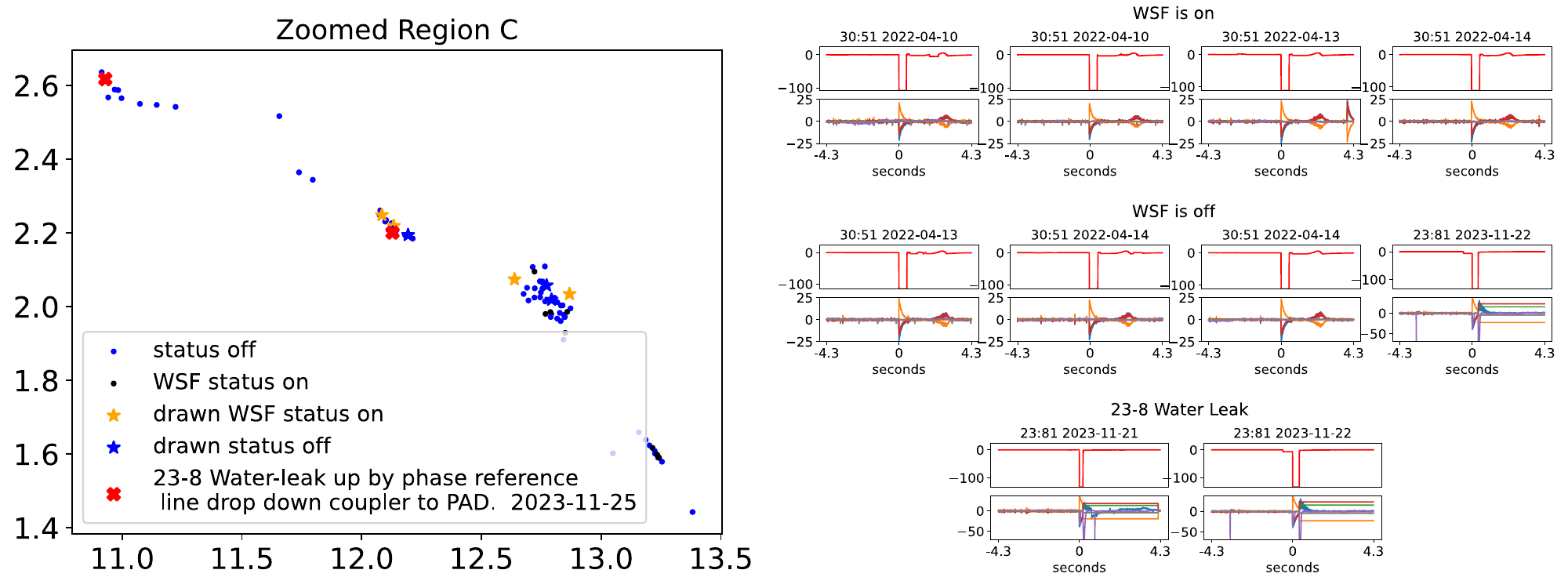}
    }

    \caption{Zoomed view of \cref{fig:clustering_status} (at left) along with phase and BPM data for select examples (at right). Each example consists of two components: the anomalous phase signals (shown in the first row) and the corresponding BPM signals (displayed in the second row).  In zoomed region A, 85\% of the examples have EVO status bits as ON. In zoomed region B, 8\% and 2\% of RE status bits and EVC status bits respectively are ON, highlighting the poor recall of some status bits. EVC and RE examples are plotted together given the similarity of phase data in examples corresponding to those status bits. In zoomed region C, 16\% of the WSF status bits are ON. }
    \label{fig:zoomed_view_cluster_status}
\end{figure*}

\section{CANDIDATE SELECTION} 

\label{sec:most_anomalous_rf_station}

This section outlines the procedure for selecting the most anomalous RF stations as part of the candidate generation step described in \cref{sec:anomaly_candidate_identification}. Note that this method is specifically tailored to our dataset and includes certain specific preprocessing steps.

To identify the most anomalous klystron stations, we begin by computing the absolute phase values for each RF station at every timestamp. Let \( \Phi \in \mathbb{R}^{N \times T} \) denote the phase matrix, where \( N \) is the number of RF stations and \( T \) is the total number of time steps. At each time \( t \in \{1, \dots, T\} \), the absolute phase for station \( n \in \{1, \dots, N\} \) is computed as
\[
\phi_{n,t} = \left| \Phi_{n,t} \right|,
\]
where \( \Phi_{n,t} \) represents the original (signed) phase value. The absolute value is used because we are interested in the magnitude of deviation from zero, regardless of sign.

To reduce the inclusion of system-level faults—characterized by uniform increases in phase values across multiple stations—we apply a preprocessing step that subtracts the fifth-largest absolute phase value at each time point. Specifically, let \( \phi_t^{(5)} \) denote the fifth-largest value among \( \{\phi_{1,t}, \dots, \phi_{N,t}\} \). The adjusted phase for station \( n \) at time \( t \) is then given by
\[
\tilde{\phi}_{n,t} = \phi_{n,t} - \phi_t^{(5)}.
\]

Next, for each station \( n \), we define a station-specific time window \( (t_i^n - k, t_i^n] \), where \( t_i^n \) is the trigger time associated with station \( n \), and \( k \) is the number of preceding pulses to consider. This window is intended to capture the earliest onset of anomalous behavior. Within this window, we compute the maximum adjusted phase:
\[
d_n = \max_{t \in (t_i^n - k, \, t_i^n]} \tilde{\phi}_{n,t}.
\]

We then identify the five stations with the largest values of \( d_n \) and designate them as anomaly candidates. From this set, we retain only those stations for which \( d_n > \tau \), where \( \tau = 25 \) is a predefined threshold. If none of the five candidates exceeds this threshold, we instead select the single station with the largest \( d_n \) value.

If only one station is selected, its RF signal is paired with the corresponding beam data to form a single candidate instance. If multiple stations exceed the threshold, each selected RF signal is paired with the same beam data, resulting in multiple candidate instances for further analysis.
Formally, let us denote the RF diagnostic data as \( s \), describing the subsystem, and the BPM data as \( q \), representing the overall quality of the system. Consequently, each candidate \( C_i \) is defined as \( C_i = \{s_i, q_i\} \). The method of identifying the most anomalous klystron station(s) is summarized in \textbf{Algorithm 1}. 

\begin{algorithm}[htbp]
\caption{Most Anomalous Klystron Station Identification (Station-Specific Trigger Times)}
\KwIn{Phase matrix \( \Phi \in \mathbb{R}^{N \times T} \), trigger times \( \{ t_i^n \}_{n=1}^N \), window size \( k \), threshold \( \tau = 25 \)}
\KwOut{Set of selected anomalous RF stations \( \mathcal{A} \)}

\For{each time \( t = 1 \) to \( T \)}{
    Compute absolute phases: \( \phi_{n,t} \gets |\Phi_{n,t}| \) for all \( n \in [1, N] \)\;
    Find the 5th-largest phase across all \( n \): \( \phi^{(5)}_t \)\;
    \For{each station \( n = 1 \) to \( N \)}{
        Adjusted phase: \( \tilde{\phi}_{n,t} \gets \phi_{n,t} - \phi^{(5)}_t \)\;
    }
}

\For{each station \( n = 1 \) to \( N \)}{
    Define station-specific window: \( \mathcal{W}_n \gets \{ \tilde{\phi}_{n,t} \mid t \in (t_i^n - k, t_i^n] \} \)\;
    Compute max deviation: \( d_n \gets \max(\mathcal{W}_n) \)\;
}

\( \mathcal{C} \gets \text{indices of top 5 stations with largest } d_n \)\;
\( \mathcal{A} \gets \{ n \in \mathcal{C} \mid d_n > \tau \} \)\;

\If{\( \mathcal{A} = \emptyset \)}{
    \( n^* \gets \arg\max_{n \in \mathcal{C}} d_n \)\;
    \( \mathcal{A} \gets \{ n^* \} \)\;
}

\Return \( \mathcal{A} \)\;

\end{algorithm}

\bibliographystyle{unsrt}
\bibliography{references}

\end{document}